\documentclass[aps,prl,twocolumn,superscriptaddress,showpacs,preprintnumbers,amsmath,amssymb]{revtex4}
\usepackage{graphicx} 
\usepackage{dcolumn}  

\graphicspath{{ps}}

\newcommand{\ee}{e^{+}e^{-}}

\newcommand{\uubar}{u\bar{u}}
\newcommand{\ddbar}{d\bar{d}}
\newcommand{\ssbar}{s\bar{s}}
\newcommand{\qqbar}{q\bar{q}}

\newcommand{\pim}{\pi^{-}}
\newcommand{\pip}{\pi^{+}}

\newcommand{\kpl}{K^{+}}

\newcommand{\rt}{\rightarrow}

\newcommand{\etal}{\em et al.}

\newcommand{\yones}{\Upsilon(1S)}
\newcommand{\ytwos}{\Upsilon(2S)}
\newcommand{\yonetwos}{\Upsilon(1,2S)}
\newcommand{\yns}{\Upsilon(nS)}

\newcommand{\lm}{\Lambda}

\newcommand{\lmb}{\bar{\Lambda}}
\newcommand{\lmlm}{\Lambda\Lambda}

\newcommand{\lmppi}{\Lambda p \pi^-}
\newcommand{\lmbpbpi}{\bar{\Lambda} \bar{p} \pi^+}

\newcommand{\xim}{\Xi^{-}}

\begin{document}



\title{ \quad\\[0.5cm] Search for an $H$-dibaryon with mass near $2m_{\lm}$ in  $\yones$ and $\ytwos$ decays}



\noaffiliation
\affiliation{University of the Basque Country UPV/EHU, 48080 Bilbao}
\affiliation{Budker Institute of Nuclear Physics SB RAS and Novosibirsk State University, Novosibirsk 630090}
\affiliation{Faculty of Mathematics and Physics, Charles University, 121 16 Prague}
\affiliation{University of Cincinnati, Cincinnati, Ohio 45221}
\affiliation{Justus-Liebig-Universit\"at Gie\ss{}en, 35392 Gie\ss{}en}
\affiliation{Gifu University, Gifu 501-1193}
\affiliation{Gyeongsang National University, Chinju 660-701}
\affiliation{Hanyang University, Seoul 133-791}
\affiliation{University of Hawaii, Honolulu, Hawaii 96822}
\affiliation{High Energy Accelerator Research Organization (KEK), Tsukuba 305-0801}
\affiliation{Ikerbasque, 48011 Bilbao}
\affiliation{Indian Institute of Technology Guwahati, Assam 781039}
\affiliation{Indian Institute of Technology Madras, Chennai 600036}
\affiliation{Institute of High Energy Physics, Chinese Academy of Sciences, Beijing 100049}
\affiliation{Institute of High Energy Physics, Vienna 1050}
\affiliation{Institute for High Energy Physics, Protvino 142281}
\affiliation{INFN - Sezione di Torino, 10125 Torino}
\affiliation{Institute for Theoretical and Experimental Physics, Moscow 117218}
\affiliation{J. Stefan Institute, 1000 Ljubljana}
\affiliation{Kanagawa University, Yokohama 221-8686}
\affiliation{Institut f\"ur Experimentelle Kernphysik, Karlsruher Institut f\"ur Technologie, 76131 Karlsruhe}
\affiliation{Korea Institute of Science and Technology Information, Daejeon 305-806}
\affiliation{Korea University, Seoul 136-713}
\affiliation{Kyungpook National University, Daegu 702-701}
\affiliation{\'Ecole Polytechnique F\'ed\'erale de Lausanne (EPFL), Lausanne 1015}
\affiliation{Faculty of Mathematics and Physics, University of Ljubljana, 1000 Ljubljana}
\affiliation{University of Maribor, 2000 Maribor}
\affiliation{Max-Planck-Institut f\"ur Physik, 80805 M\"unchen}
\affiliation{School of Physics, University of Melbourne, Victoria 3010}
\affiliation{Moscow Physical Engineering Institute, Moscow 115409}
\affiliation{Moscow Institute of Physics and Technology, Moscow Region 141700}
\affiliation{Graduate School of Science, Nagoya University, Nagoya 464-8602}
\affiliation{Kobayashi-Maskawa Institute, Nagoya University, Nagoya 464-8602}
\affiliation{Nara Women's University, Nara 630-8506}
\affiliation{National Central University, Chung-li 32054}
\affiliation{National United University, Miao Li 36003}
\affiliation{Department of Physics, National Taiwan University, Taipei 10617}
\affiliation{H. Niewodniczanski Institute of Nuclear Physics, Krakow 31-342}
\affiliation{Nippon Dental University, Niigata 951-8580}
\affiliation{Niigata University, Niigata 950-2181}
\affiliation{University of Nova Gorica, 5000 Nova Gorica}
\affiliation{Osaka City University, Osaka 558-8585}
\affiliation{Pacific Northwest National Laboratory, Richland, Washington 99352}
\affiliation{University of Pittsburgh, Pittsburgh, Pennsylvania 15260}
\affiliation{Research Center for Electron Photon Science, Tohoku University, Sendai 980-8578}
\affiliation{Seoul National University, Seoul 151-742}
\affiliation{Sungkyunkwan University, Suwon 440-746}
\affiliation{Tata Institute of Fundamental Research, Mumbai 400005}
\affiliation{Excellence Cluster Universe, Technische Universit\"at M\"unchen, 85748 Garching}
\affiliation{Tohoku Gakuin University, Tagajo 985-8537}
\affiliation{Tohoku University, Sendai 980-8578}
\affiliation{Department of Physics, University of Tokyo, Tokyo 113-0033}
\affiliation{Tokyo Institute of Technology, Tokyo 152-8550}
\affiliation{Tokyo Metropolitan University, Tokyo 192-0397}
\affiliation{CNP, Virginia Polytechnic Institute and State University, Blacksburg, Virginia 24061}
\affiliation{Wayne State University, Detroit, Michigan 48202}
\affiliation{Yamagata University, Yamagata 990-8560}
\affiliation{Yonsei University, Seoul 120-749}
  \author{B.~H.~Kim}\affiliation{Seoul National University, Seoul 151-742} 
  \author{S.~L.~Olsen}\affiliation{Seoul National University, Seoul 151-742} 
  \author{I.~Adachi}\affiliation{High Energy Accelerator Research Organization (KEK), Tsukuba 305-0801} 
  \author{H.~Aihara}\affiliation{Department of Physics, University of Tokyo, Tokyo 113-0033} 
  \author{D.~M.~Asner}\affiliation{Pacific Northwest National Laboratory, Richland, Washington 99352} 
  \author{V.~Aulchenko}\affiliation{Budker Institute of Nuclear Physics SB RAS and Novosibirsk State University, Novosibirsk 630090} 
  \author{A.~Bay}\affiliation{\'Ecole Polytechnique F\'ed\'erale de Lausanne (EPFL), Lausanne 1015} 
  \author{K.~Belous}\affiliation{Institute for High Energy Physics, Protvino 142281} 
  \author{B.~Bhuyan}\affiliation{Indian Institute of Technology Guwahati, Assam 781039} 
  \author{G.~Bonvicini}\affiliation{Wayne State University, Detroit, Michigan 48202} 
% ...
\collaboration{The Belle Collaboration}

\begin{abstract}

We report the results of a high-statistics search for 
$H$-dibaryon production in inclusive $\yones$ and $\ytwos$ 
decays.  No indication of an $H$-dibaryon with mass
near the $M_H=2m_{\lm}$ threshold
is seen in either the $H\rt \lmppi$ or $\lmlm$ decay
channels and 90\% confidence level
branching-fraction upper limits are set that are
between one and two orders of magnitude below the
measured branching fractions for inclusive $\yones$
and $\ytwos$ decays to antideuterons.  Since
$\yonetwos$ decays produce flavor-$SU(3)$-symmetric
final states, these results put stringent constraints 
on $H$-dibaryon properties.
The results are based on analyses of
102~million $\yones$ and 158~million $\ytwos$ events
collected with the Belle detector at the KEKB $\ee$ collider.

\end{abstract}


\maketitle


{\renewcommand{\thefootnote}{\fnsymbol{footnote}}}
\setcounter{footnote}{0}


In 1977, Jaffe
predicted the existence of a doubly strange,
six-quark structure ($uuddss$) with quantum numbers
$I=0$ and $J^P=0^+$ and a mass that is $\simeq 80$~MeV
below the $2m_{\lm}$ threshold, which he dubbed the
$H$-dibaryon~\cite{jaffe_H}.  
An $S=-2$, baryon-number $B=2$
particle with mass below $2m_{\lm}$ would decay via weak
interactions and, thus, be long-lived with a lifetime comparable
to that of the $\lm$ and negligible natural width.  

Jaffe's specific prediction was ruled out by the observation
of double-$\lm$ hypernuclei events~\cite{double-lambda,nagara,e176},  
especially the famous ``Nagara'' event that has a relatively unambiguous
signature as a $_{\lmlm}^6$He hypernucleus produced via
$\xim$ capture in emulsion~\cite{nagara}.
The measured $\lmlm$ binding energy,
$B_{\lmlm}=7.13\pm 0.87$~MeV, establishes, with a 90\% confidence level
(CL), a lower limit of $M_{H}>2223.7$~MeV, severely narrowing the window
for a stable $H$ to the binding energy range
$B_H\equiv 2m_{\Lambda}-M_H < 7.9$~MeV.

Although Jaffe's original prediction for $B_H\simeq 81$ MeV
has been ruled out,  the theoretical case for
an $H$-dibaryon with a mass near $2m_{\lm}$ continues to be
strong and has been recently strengthened by lattice QCD
calculations (LQCD) by the NPLQCD~\cite{NPLQCD,NPLQCD_2} and
HALQCD~\cite{HALQCD} collaborations that both find 
a bound $H$-dibaryon, albeit for non-physical values for
the pion mass.  
NPLQCD's linear (quadratic) extrapolation to
the physical pion mass gives $B_H= -0.2\pm 8.0$~MeV
($7.4\pm 6.2$~MeV)~\cite{NPLQCD_2}.  Carames and
Valcarce~\cite{Carames} recently studied the $H$ with a
chiral constituent model constrained by $\Lambda N$,
$\Sigma N$, $\Xi N$ and $\lm\lm$ cross section data
and find  $B_H$ values that are similar to the NPLQCD
extrapolated values. 

These recent theoretical results motivate searches for the $H$
with mass near the $M_H = 2m_{\Lambda}$ threshold. For masses below
threshold, the $H$ would predominantly decay via $\Delta S=+1$
weak transitions to $\Lambda n$, $\Sigma^- p$, $\Sigma^0 n$ or
$\lmppi$ final states.  For masses above $2m_{\lm}$, but below 
$m_{\Xi^0} + m_n~(=2m_{\lm}+ 23.1$~MeV), the $H$ would decay
via strong interactions to $\lmlm$ 100\% of the time.
The E522 collaboration at KEK studied
$\lm\lm$ production in the $^{12}C(K^-,K^+\lm\lm X)$
reaction and reported an intriguing near-threshold
enhancement but with limited statistics~\cite{e522}. 
The BNL-E836 collaboration searched for the
$\Delta S=+2$ reaction $^3{\rm He}(K^-,K^+)Hn$ and established
cross section limits spanning the range
$50~{\rm MeV}\le B_H \le 380~{\rm MeV}$~\cite{e836}.  Searches for a
bound $H$ decaying to $\lmppi$ reported negative
results~\cite{KTeV,e910}.  Earlier searches, also with negative
results, are listed in Ref.~\cite{H-searches}.

Decays of narrow $\yns$ ($n=1,2,3$) bottomonium
($b\bar{b}$) resonances are
particularly well suited for searches for multiquark states with
non-zero strangeness.  The $\yns$ states are flavor-$SU(3)$
singlets and primarily decay via the 
three-gluon annihilation process  ({\it e.g.}, 
(${\mathcal B}(\yones\rt ggg) = 81.7\pm 0.7$\%~\cite{PDG}). 
The gluons materialize into $\uubar$, $\ddbar$ and $\ssbar$
pairs in roughly equal numbers. 
The high density of quarks and antiquarks in the limited final-state phase
space is conducive to the production of
multi-quark systems, as demonstrated by large branching
fractions for inclusive antideuteron ($\bar{D}$) production:
${\mathcal {B}}(\yones\rt \bar{D}\, X)=
(2.9\pm 0.3)\times 10^{-5}$ and 
${\mathcal {B}}(\ytwos\rt \bar{D}\, X)=
(3.4\pm 0.6)\times 10^{-5}$~\cite{CLEO_dbar}.  An upper limit for
the production of a six-quark $S=-2$ state in $\yns$ decays that is
substantially below that for the six-quark antideuteron would
be strong evidence against its existence.

Here we report results of a search for
$H$-dibaryon production in the inclusive processes 
$\yonetwos \rt H \, X$; $H\rt \lmppi$ and $\lmlm$~\cite{conj}.
We use data samples containing 102 million $\yones$ and
158 million $\ytwos$ decays collected with the Belle detector
operating at the KEKB $\ee$ collider~\cite{KEKB}.
The data were accumulated at center-of-mass system (cms)
energies of $\sqrt{s} = 9.460$~GeV and $10.023$~GeV, which correspond
to the $\yones$ and $\ytwos$ resonance peaks, respectively.  
Contributions from the $\ee\rt\qqbar$ 
($q=u,\, d,\, s,\,\mathrm{and}\, c$) continuum process are inferred from
a 63.7~fb$^{-1}$ sample collected
at $\sqrt{s} = 10.53$~GeV and scaled by luminosity and $1/s$.
We assume equal $\yones$ and $\ytwos$ branching fractions:
{\it i.e.,} ${\mathcal {B}}(\yones\rt H\, X)={\mathcal {B}}(\ytwos\rt H\, X)
\equiv {\mathcal {B}}(\yonetwos\rt H\, X)$.

Belle is a large-solid-angle magnetic 
spectrometer consisting of a silicon vertex 
detector, a cylindrical drift chamber (CDC), an 
array of aerogel threshold Cherenkov counters (ACC),  a 
barrel-like arrangement of time-of-flight  scintillation 
counters (TOF), and an electromagnetic calorimeter
(ECL) comprised of CsI(Tl) crystals  located inside
a superconducting solenoid coil that provides a 1.5~T
magnetic field.  
Measurements of $dE/dx$ in the
CDC, ACC light yields, TOF flight times and ECL
energy deposits 
are combined to form particle identification (pid) likelihoods 
${\mathcal {L}}(h)$
($h=e^+, \pip,\,\kpl \, \mathrm{or}\, p)$ for charged 
tracks.  The
${\mathcal R}(h|h') =
{\mathcal {L}(h)}/({\mathcal {L}}(h)+ {\mathcal {L}}(h'))$
ratios are used to make pid assignments.
Belle is described in detail elsewhere~\cite{Belle}.

Samples of simulated $\yones$ and $\ytwos$ Monte Carlo (MC) events,
generated with PYTHIA~\cite{pythia} and simulated using GEANT3~\cite{geant3},
are used to study backgrounds and determine
efficiencies.   For signal MC for various $H$ decay modes, we use
PYTHIA with the $\Xi^{*0}(1530)$ mass, width and decay-table entries
replaced with hypothesized parameters for the $H$.  For MC-based
optimization of selection criteria, we optimize a figure of merit defined
as $FoM = n_{\rm sig}/\sqrt{n_{\rm sig} + n_{\rm bkg}}$, where $n_{\rm sig}$
($n_{\rm bkg}$) is the number of selected signal (background) events
assuming 
${\mathcal {B}}(\yns\rt H\, X)=3 \times 10^{-5}$.

For both investigated channels, event selection starts
with the identification of a $\lm$ candidate reconstructed via its
$p\pim$ decay using the $\lm$-momentum-dependent criteria
based on proton pid, track vertex information, decay length,
and $M(p\pim)$ described in Ref.~\cite{belle_lambda}. The $M(p\pim)$
distribution for selected candidates is well fitted by a Lorentzian
function with a FWHM resolution for the $\lm$ peak of $1.50\pm 0.01$~MeV. 
For $\lm$ candidates, we require $\Delta M_{\lm}\equiv |M(p\pi)-m_{\lm}|<3.0$~MeV. 

For the $H\rt\lmppi$ search, the $p\pim$ track selection requirements
are optimized using $FoM$s determined by MC assuming $\tau_H=\tau_{\lm}$.
Both the $p$ and $\pim$ are required to be well identified by the
pid measurements:   ${\mathcal R}(p|h^+)>0.9$ ($h^+=\pip$ or $K^+$) ${\mathcal R}(\pim|e^-)>0.9$ 
and ${\mathcal R}(\pim|K^-)>0.6$~\cite{pideff_lppi}.  
We require that the $p$ and $\pi^-$ tracks and the $\lm$
trajectory satisfy a fit to a common vertex with $\chi^2_{\lmppi}\le 50$.  In addition we 
require $c\tau_{\lmppi}\ge 0.0$, where
$c\tau\equiv\vec{\ell}\cdot\vec{p}_H M_{H}/|\vec{p}_H|^{2}$ and $\vec{\ell}$ is the
displacement between the run-dependent average interaction point (IP)
and the fitted vertex position. 
In some cases, the tracking algorithm finds two reconstructed tracks with
nearly the same parameters from CDC hits produced by a single particle.
Contamination from this source is removed by the requirements $M(p_1 p_2)\ge 1878$~MeV,
$M(\pim_1 \pim_2)\ge 280$~MeV and $N_{\rm hits}(p_1)+N_{\rm hits}(p_2)\ge 50$, where
$H\rt\lm p_2\pim_2$; $\lm\rt p_1\pim_1$ and $N_{\rm hits}(p_i)$ is the number of CDC hits
used to reconstruct the $i^{\rm th}$ proton.  In the $\lmppi$ mode, there is a large background
from $\lm$ and $p$ production via secondary interactions in the material of
the beampipe and inner detector.   This is removed by requiring $|\vec{p}_{h}|>0.5$~GeV
for both $h=\lm$ and $h=p$; this requirement is not applied to the $\lmbpbpi$ channel.
In 6.3\% (5.2\%) of the data (MC) events, there are two or more entries that have one or
more tracks in common.  In these cases, the combination with the smallest $\chi^2_{\lmppi}$
value is selected.  For signal MC events, this chooses the correct combination 93.4\%
of the time.  The $\lm\rt p_1\pim_1$ candidate is subjected to a kinematic fit that
constrains $M(p_1\pim_1)$ to $m_{\lm}$.  The final selection efficiencies are
determined from MC by averaging $\yones$ \& $\ytwos$ signal MC to be
$\epsilon_1= 7.7\%$ for $H\rt\lmppi$ and $\bar{\epsilon}_1 =8.8\%$ for $\bar{H}\rt\lmbpbpi$.

\begin{figure}[htb]
\begin{center}
\includegraphics[height=0.25\textwidth,width=0.5\textwidth]
{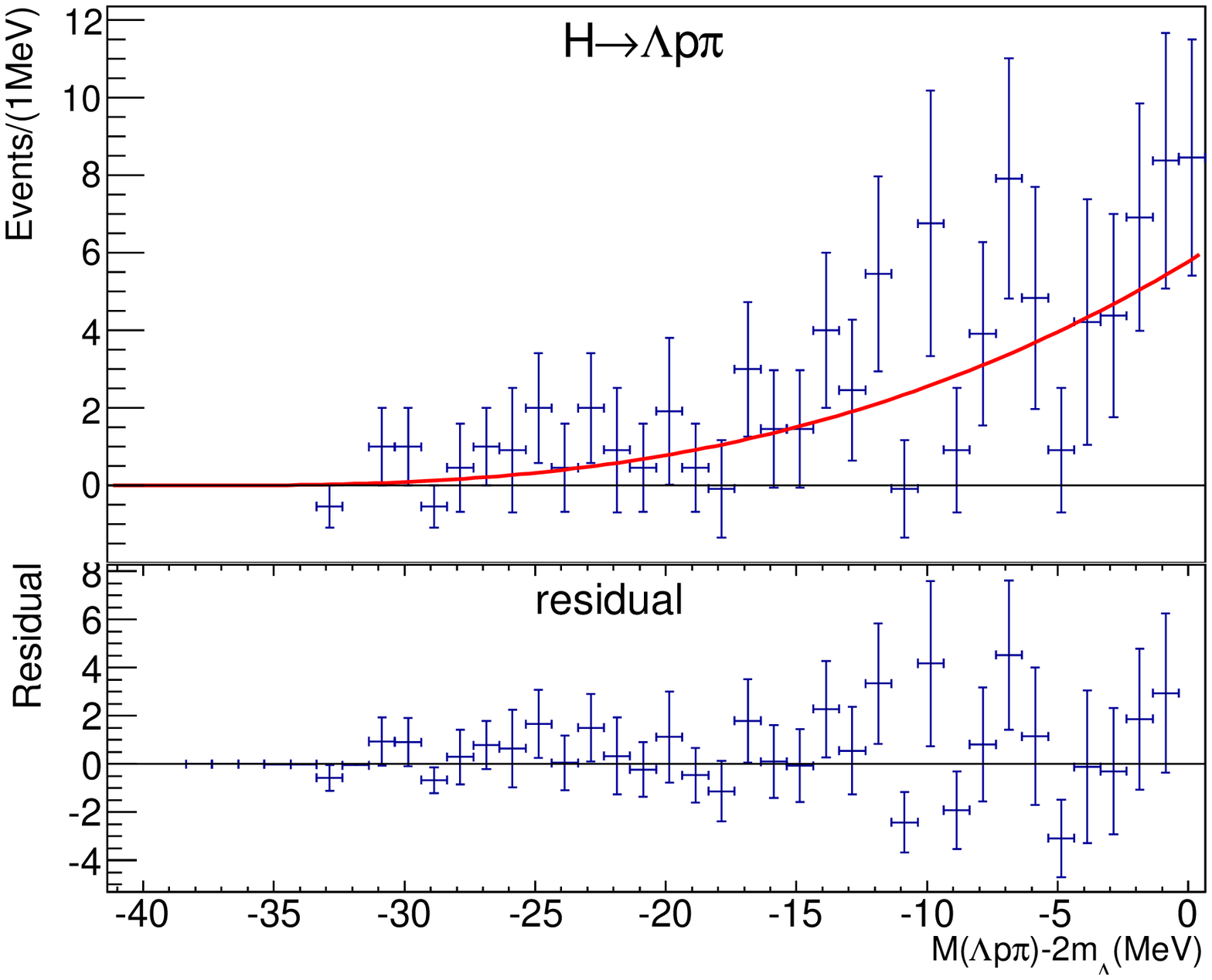}
\end{center}
\begin{center}
\includegraphics[height=0.25\textwidth,width=0.5\textwidth]
{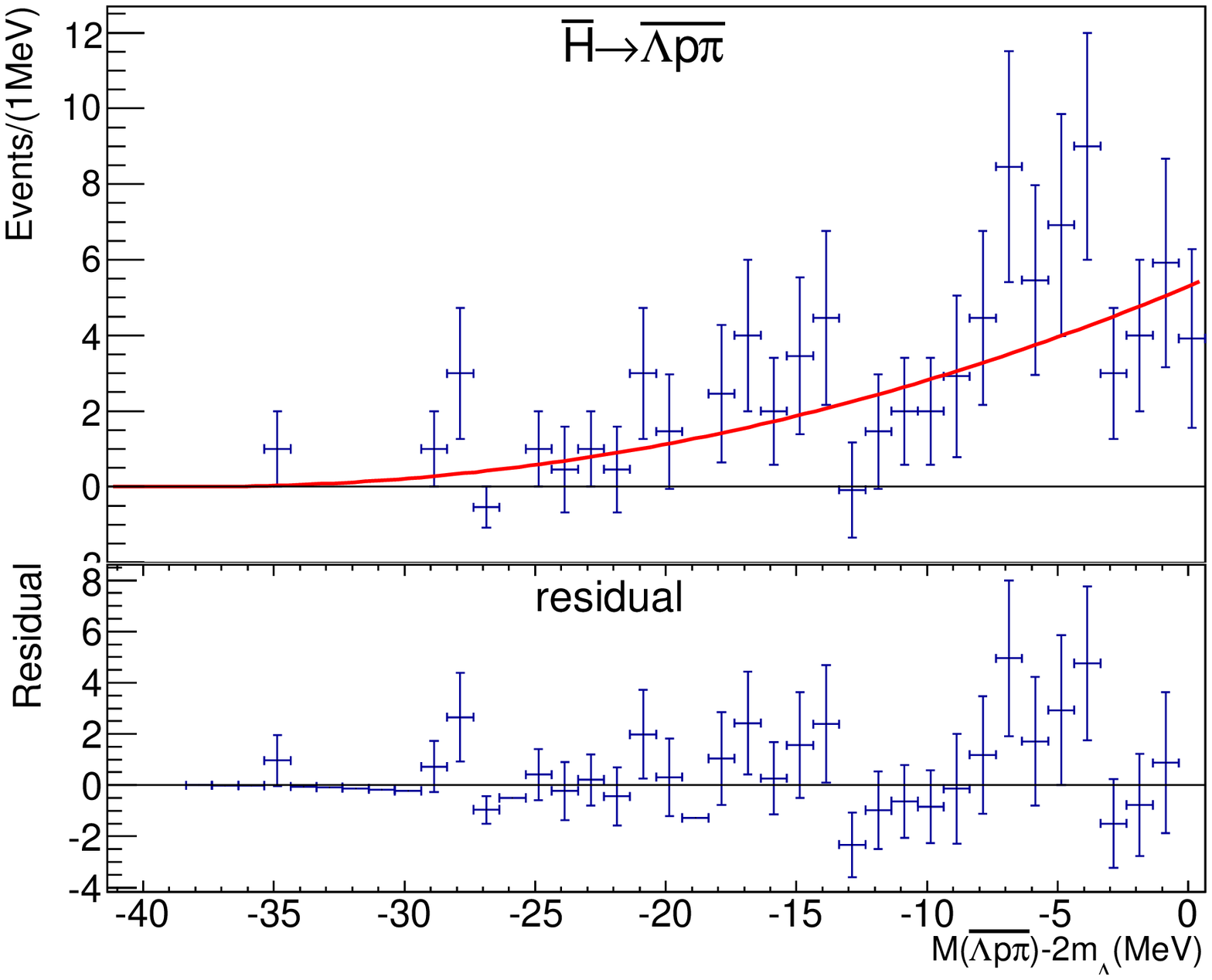}
\end{center}

\caption{ 
{\bf Top:}  The continuum-subtracted $M(\lmppi)$ distribution (upper) and fit residuals (lower)
for the combined $\yones$ and $\ytwos$ data samples.  The curve shows the
results of the background-only fit described in the text.
{\bf Bottom:} The corresponding $M(\lmbpbpi ) $ distributions.
}
\label{fig:data-lppi}
\end{figure}

The resulting continuum-subtracted $M(\lmppi)$ ($M(\lmbpbpi)$)
distribution for the combined $\yones$ and $\ytwos$ samples,
shown in the top (bottom) panel of Fig.~\ref{fig:data-lppi}, has no evident 
$H\rt\lmppi$ ($\bar{H}\rt\lmbpbpi$) signal.  The curve in the figure is the result of a fit using 
an ARGUS-like threshold function to model the background~\cite{ARGUS}; fit residuals are also shown.
 
For the second $\Lambda$ ($\lm_2$) in the $H\rt\lm_1\lm_2$
($\lm_i\rt p_i\pim_i$) channel, in addition to the criteria
used for $\lm_1$ selection, $FoM$s based on MC events are used to
optimize the additional requirements $\chi^2_{\lm_1\lm_2}<200$ from a
$\lm_1\lm_2$ vertex and IP constrained fit, and $c\tau_{\lm_2}\ge -0.5$~cm.
Entries in which two of the selected tracks originate from a single particle
are removed by the requirements $M(p_1 p_2)\ge 1878$~MeV,
$M(\pim_1 \pim_2)\ge 288$~MeV and  $N_{\rm hits}(p_1)+N_{\rm hits}(p_2)\ge 60$.
In 3.2\% (2.8\%) of the data (MC) events, two or more entries have one
or more tracks in common.  In these cases, we choose the track combination
that has the smallest $\chi^2_{\lm_1\lm_2}$ value.  For signal MC events, this
selects the correct combination 95.4\% of the time.  The $\lm\lm$ candidates
are subjected to a kinematic fit that constrains both $p\pim$ masses to $m_{\lm}.$
The MC-determined selection efficiencies, obtained by averaging $\yones$ \& $\ytwos$
signal MC results, are $\epsilon_2= 10.9\%$
for $H\rt\lm\lm$ and $\bar{\epsilon}_2 =10.1\%$ for $\bar{H}\rt\lmb\lmb$.

\begin{figure}[htb]
\begin{center}
\includegraphics[height=0.25\textwidth,width=0.5\textwidth]
{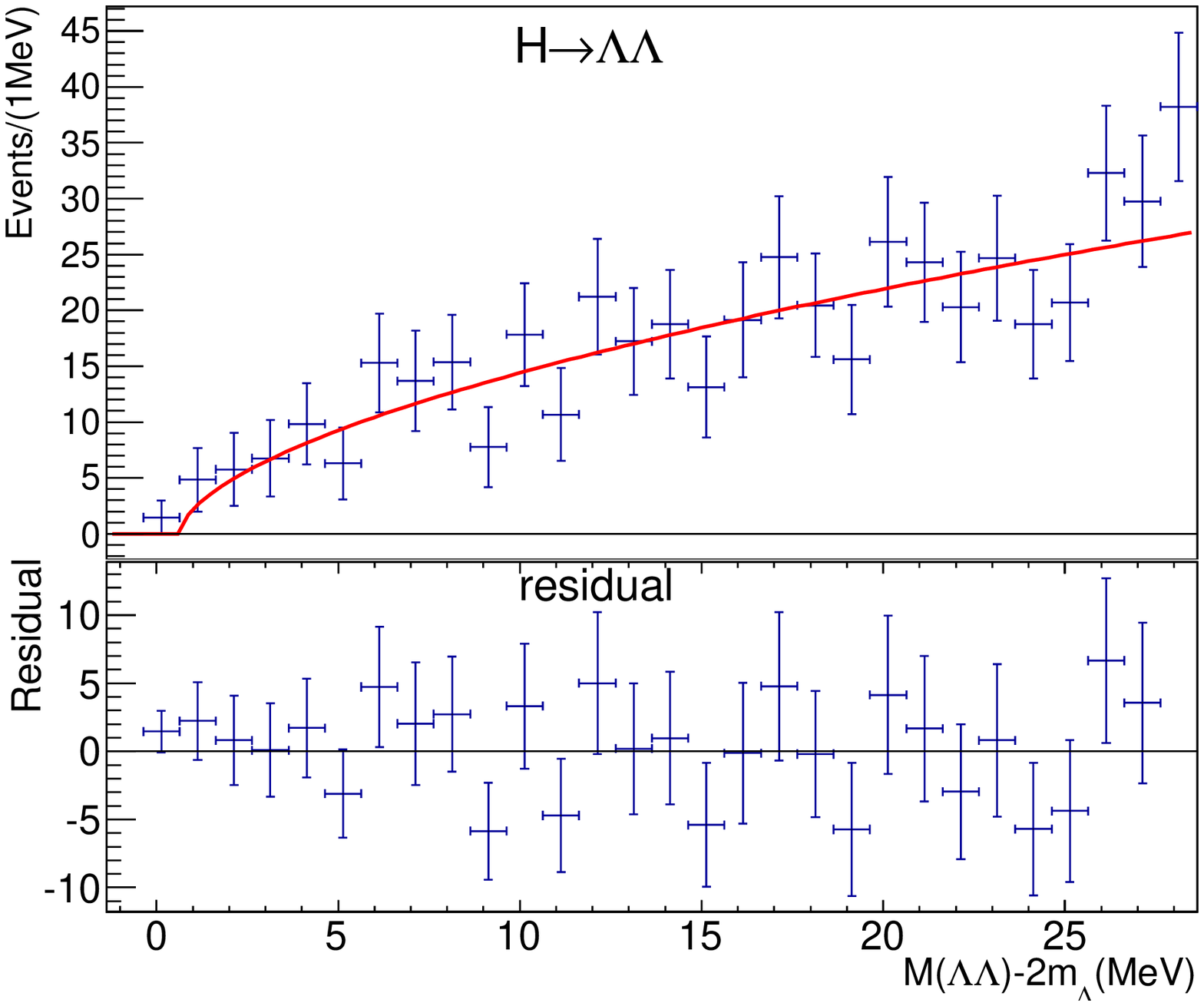}
\end{center}
\begin{center}
\includegraphics[height=0.25\textwidth,width=0.5\textwidth]
{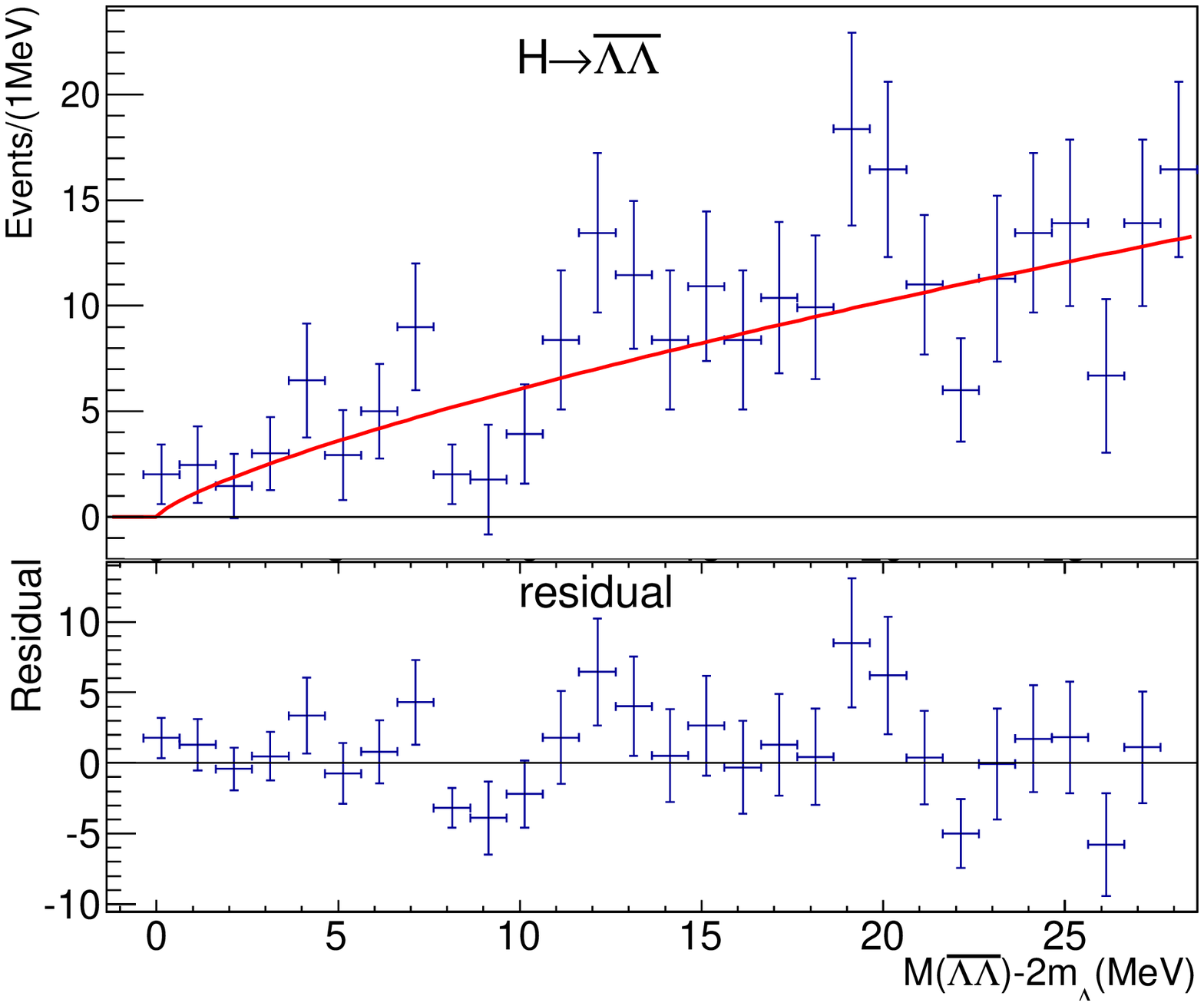}
\end{center}
\caption{ 
{\bf Top:} The continuum-subtracted $M(\lm\lm)$ distribution (upper)
and fit residuals (lower) for the
combined $\yones$ and $\ytwos$ data samples with the 
background-only fit superimposed.
{\bf Bottom:} The corresponding
$M(\lmb\lmb)$ distributions.
}
\label{fig:data-ll}
\end{figure}
  
The difference between the $\lm\lm$ and $\lmb\lmb$ 
signal yields in the region $M(\lm\lm)<2.38$~GeV,
determined from two-dimensional fits to scatter plots
of $M(p_1\pi_1)$ {\it vs.} $M(p_2\pi_2)$ with the
$\lm$ mass requirements relaxed, is larger than the
difference in the MC-determined $\lm\lm$
and $\lmb\lmb$ acceptances.  This
is attributed to deficiencies in the simulation of low-energy
$\lmb$ and $\bar{p}$ inelastic interactions in the material
of the inner detector.  To account for this, a correction factor
of $0.83\pm 0.13$ is applied to the $H\rt\lmb\lmb$ and
$H\rt\lmb \bar{p}\pip$
efficiencies.  The error on this factor is included in the
systematic error.

The continuum-subtracted $M(\lm\lm)$ ($M(\lmb\lmb)$)
distribution for events that satify all of the selection requirements
is shown in the top (bottom) panel of Fig.~\ref{fig:data-ll}, where there
is no sign of a near-threshold enhancement similar to that reported by the
E522 collaboration~\cite{e522} nor any other evident signal for
$H\rt\lm\lm$ ($\bar{H}\rt\lmb\lmb$).  The curve is the result of a
background-only fit using the functional form described above; fit
residuals are also shown.
 
For each channel, we do a sequence of binned 
fits to the invariant mass distributions in Figs.~\ref{fig:data-lppi}
and \ref{fig:data-ll} using a signal function to represent
$H\rt f_i$ ($f_1= \lmppi$ \& $f_2 = \lm\lm$)
and an ARGUS function to represent the background.  In the
fits, the signal peak position is confined to a 4~MeV window that is scanned in
4~MeV steps across the ranges $(m_{\lm}+m_{p}+m_{\pim})\le M(\lmppi)\le 2m_{\lm}$ and
$2m_{\lm}\le M(\lm\lm)\le 2m_{\lm}+28$~MeV.  For the $\lmppi$ ($\lmbpbpi$) mode, the signal function
is a Gaussian whose resolution width is fixed at its MC-determined value
scaled by a factor $f=0.85 (1.12)$ that is determined from a comparison of data
and MC fits to inclusive $\xim\rt\lm\pi^-$ and $\Xi^0_c(2470)\rt\xim\pi^+$ signals
found in the same data samples.  For the $\lm\lm$
mode, the signal function is a Lorentzian with FWHM fixed at either 
$\Gamma= 0~{\rm or}~10$~MeV convolved with a Gaussian.  Since
the $f_i$ and $\bar{f_i}$ acceptances are different, we fit
the particle and antiparticle distributions separately.

None of the fits exhibit a positive signal with greater than $3\sigma$ significance.
The fit results are translated into 90\% CL upper limits on the signal yield,
$ N_{i}^{\rm UL}(M_H)$ and $\bar{N}_{i}^{UL}(M_H)$, by convolving the fit likelihood
distribution with a Gaussian whose width equals the systematic error
(discussed below) and then determining the yield below which 90\% of the area
above $N_i=0$ is contained.  These values are used to  
determine upper limits on the inclusive product branching fractions
via the relation
\begin{eqnarray}
\nonumber
{\mathcal B}(\Upsilon(1,2S)\rt H~X)\cdot {\mathcal B}(H \rt f_i) < ~~~~~~~~~~\\
\frac{1}
{2N_{\Upsilon}({\mathcal B}_{\lm\rt p\pim})^{i}}
\frac{N^{\rm UL}_{i}(M_H)}{\epsilon_i},
\end{eqnarray}
where $N_{\Upsilon} = (260 \pm 6)\times 10^{6}$ is the total number of
$\yones$ plus $\ytwos$ events in the data sample~\cite{nyns} and 
${\mathcal B}_{\lm\rt p\pim}=0.639\pm 0.005$~\cite{PDG}.  

Sources of systematic errors and their contributions are listed in
Table~\ref{tbl:syst}.   The tracking, pid and $\lm$ reconstruction
uncertainties are common to other Belle analyses and are determined
from data-MC comparisons of various control samples.  For the channel-specific
vertex requirements, we use data-MC differences found in high-statistics samples
of inclusive 
$\yonetwos\rt\lm\bar{p}\pip$
and $\lm\lmb$ events with $M(\lm\bar{p}\pi)<2.28$~GeV 
($M(\lm\lmb)<2.38$~GeV) selected with the same vertex criteria.
The continuum subtraction systematic error
contribution is determined from the errors in the relative on- and off-resonance
luminosity measurements. Systematic errors associated with the MC-determined
acceptance and minimum momentum requirement are determined
by varying parameters used in the PYTHIA generator and GEANT simulation programs.
The systematic errors associated with the signal fitting are determined from
changes induced by variations in the binning and fitting ranges in fits to an inclusive
$\Xi^0_c(2470)\rt\xim\pip$ signal seen in the same data sample.  Quadratic sums 
of the individual contributions are taken as the total systematic errors. 

\begin{table}[htb]
\begin{center}
\caption{\label{tbl:syst}
Systematic error sources (in percent).  When the $H$
and $\bar{H}$ values differ, the $\bar{H}$ values are given in parentheses.
}
\begin{tabular}{l|c|c}\hline  
Source                     &~~$H\rt \lmppi$~~&~~$H \rt \lm\lm$~~    \\\hline\hline
$N_{\yones}+N_{\ytwos}$        &   $2.3$         &           $2.3$      \\
tracking                 &   $3.6$         &          $3.6$       \\
particle id               &   $7.2$         &          $4.3$       \\
$\lm$ reconstruction       &    $3.2~(5.3)$   &        $12.6~(9.6)$  \\
Vertex requirements        &   $3.9$         &           $3.5$      \\
Signal efficiency                  &    $2.0~(15.7)$   &     $1.9~(15.8)$      \\
Continuum subtraction      &   $1.4$         &           $1.4$      \\
$\mathcal{B}(\lm\rt p\pim)$&   $0.8$         &           $1.6$      \\
Fitting                    &    $2.0$         &           $2.0$     \\
Resolution                 &    $2.6$         &           $2.6$     \\

\hline
Quadrature sum             &    $10.2~(19.1)$  &       $14.7~(19.8)$   \\
\hline \hline 
\end{tabular}
\end{center}
\end{table}

\begin{figure}[htb]
\begin{center}
\includegraphics[height=0.25\textwidth,width=0.5\textwidth]
{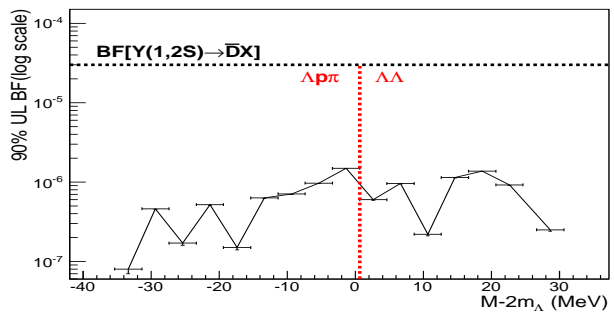}
\end{center}
\caption{ 
Upper limits (at 90\% CL) for 
${\mathcal B}(\Upsilon(1,2S)\rt H~X)\cdot {\mathcal B}(H \rt f_i)$
for a narrow ($\Gamma=0$) $H$-dibaryon {\it vs.} $M_H-2m_{\lm}$.
The vertical dotted line indicates the $M_H=2m_{\lm}$ threshold.  The
limits below (above) the $2m_{\lm}$ threshold are for $f_1= \lmppi$
($f_2=\lm\lm$). The horizontal dotted line indicates the
average PDG value for ${\mathcal B}(\Upsilon(1,2S)\rt \bar{D}~X)$.
}
\label{fig:limits}
\end{figure}

For the final limits, we use the branching fraction value
that contains $<$90\% of the above-zero area of
the product of the $H$ and $\bar{H}$ likelihood functions.    
Figure~\ref{fig:limits} shows the resulting
$M_H - 2m_{\lm}$-dependent upper limits for the $\lmppi$ and
$\lm\lm$ (for $\Gamma=0$) modes. The upper limit values,
listed in Table~\ref{tbl:limits}, are all more than an order of magnitude
lower than the average of measured values of
 ${\mathcal B}(\Upsilon(1,2S)\rt \bar{D}~X)$, shown 
in Fig.~\ref{fig:limits} as a horizontal dotted line.  

The $H\rt\lmppi$ limits quoted in Table~\ref{tbl:limits}
and shown in Fig.~\ref{fig:limits} are determined for an $H$
lifetime $\tau_{H}=0.263$~ns,  {\it i.e.,} the $\lm$ lifetime.
The acceptance decreases and, therefore, the limits 
increase, with increasing lifetime: for $\tau_{H}=5\tau_{\lm}$,
the acceptance is a factor of two lower and
the limits are correspondingly twice as high.
Conversely,  for shorter lifetimes, the acceptance increases:
for $\tau= 0.5\tau_{\lm}$, the acceptance is higher 
and the limits are more stringent by $12 \pm 2$\%.

\begin{table}[htb]
\begin{center}
\caption{\label{tbl:limits}
90\% CL upper limits ($\times 10^{-7}$) on the product branching fraction
${\mathcal B}(\Upsilon(1,2S)\rt H~X)\cdot {\mathcal B}(H \rt f_i)$,
$f_1=\lmppi$; $\delta M_1 = 2m_{\lm}- M_H$ and
$f_2=\lm\lm$; $\delta M_2 = M_H-2m_{\lm}$. 
}
\begin{tabular}{l|  c c c c c c c c c }\hline  
$\delta M_i$~(MeV)       & ~2 & ~6 & 10 & 14 & 18 & 22 & 26 & 30 & 34 \\\hline\hline
$f_1=\lmppi$ &15.~&9.7~&~7.1&~6.3&~1.5&~5.2&~1.7&~4.6&~0.8\\\hline
$f_2=\lm\lm$ &    &    &    &    &    &    &    &    & \\
~~~~~$\Gamma=0$          &6.0&~9.6&~2.2&~11.&14.&~9.2&~2.5&  \\
~~~~~$\Gamma=10$~MeV     &16.&17.&15.&37.&44.&42.&33.&   \\\hline
\end{tabular}
\end{center}
\end{table}

The results reported here are some of the most stringent constraints to date on
the existence of an $H$-dibaryon with mass near the $2m_{\lm}$ threshold~\cite{limits}.
These upper limits are between one and two orders of magnitude below the average
of the PDG value for inclusive $\yones$ and $\ytwos$ decays to antideuterons.  Since 
$\Upsilon\rt\ $hadrons decays produce final states that are flavor-$SU(3)$
symmetric, this suggests that if an $H$-dibaryon exists in this mass
range, it must have very different dynamical properties than the deuteron,
or, in the case of $M_H<2m_{\lm}$, a strongly suppressed $H\rt \lmppi$ decay mode.


We thank the KEKB group for excellent operation of the
accelerator; the KEK cryogenics group for efficient solenoid
operations; and the KEK computer group, the NII, and 
PNNL/EMSL for valuable computing and SINET4 network support.  
We acknowledge support from MEXT, JSPS and Nagoya's TLPRC (Japan);
ARC and DIISR (Australia); NSFC (China); MSMT (Czechia);
DST (India); INFN (Italy); MEST, NRF, GSDC of KISTI, and WCU (Korea); 
MNiSW (Poland); MES and RFAAE (Russia); ARRS (Slovenia); 
SNSF (Switzerland); NSC and MOE (Taiwan); and DOE and NSF (USA).
B.-H. Kim and S.~L.~Olsen acknowledge support from NRF (Korea)
Grant No. 2011-0029457 and WCU Grant No. R32-10155.


\begin{thebibliography}{99}

\bibitem{jaffe_H} R.L.~Jaffe, Phys. Rev. Lett. {\bf 38}, 195 (1977).

\bibitem{double-lambda} 
M.~Danysz, K.~Garbowska, J.~Pniewski and J.~Zakrzewski,
Phys. Rev. Lett. {\bf 11}, 29 (1963); D.J.~Prowse,
Phys. Rev. Lett. {\bf 17}, 782 (1966);
S.~Aoki, {\etal} Prog. Theor. Phys. {\bf 85}, 1287 (1991).

\bibitem{nagara} H.~Takahashi {\etal},
Phys. Rev. Lett. {\bf 87}, 212502 (2001).

\bibitem{e176} K.~Nakazawa {\etal} (E176 Collaboration),
Nucl. Phys. A {\bf 835}, 207 (2010).

\bibitem{NPLQCD} S.R.~Beane {\etal} (NPLQCD Collaboration),
Phys. Rev. Lett. {\bf 106}, 162001 (2011).

\bibitem{NPLQCD_2} S.R.~Beane {\etal} (NPLQCD Collaboration),
Mod. Phys. Lett. A {\bf 26}, 2587 (2011).

\bibitem{HALQCD} T.~Inoue {\etal} (HALQCD Collaboration),
Phys. Rev. Lett. {\bf 106}, 162002 (2011).

\bibitem{Carames} S.R.~Carames and A. Valcarce,
Rhys. Rev. C {\bf 85}, 045202 (2012).

\bibitem{e522} C.J.~Yoon {\etal} (KEK-PS E522 Collaboration),
Phys. Rev. C {\bf 75}, 022201(R) (2007).  See also
J.K.~Ahn {\etal} (KEK-PS E224 Collaboration), Phys. Lett.
{\bf B444}, 267 (1998) and J.~Belz {\etal} (BNL E888 Collaboration),
Phys. Rev. D {\bf 53}, 3487 (1996).


\bibitem{e836} R.W.~Stotzer {\etal} (BNL-E836 Collaboration),
Phys. Rev. Lett. {\bf 78}, 3646 (1997).

\bibitem{KTeV} A.~Alavi-Harati {\etal} (KTeV Collaboration),
Phys. Rev. Lett. {\bf 84}, 2593 (2003).

\bibitem{e910} I.~Chemakin {\etal} (E910 Collaboration),
Nucl. Phys. A {\bf 639}, 407c (1998).

\bibitem{H-searches} J.K.~Ahn {\etal} (KEK-PS E224 Collaboration),
Phys. Lett. B {\bf 444}, 267 (1998); J. Belz {\etal},
Phys. Rev. Lett. {\bf 76}, 3277 (1996) \& Phys. Rev. C {\bf 56}, 1164 (1997).

\bibitem{PDG} J.~Beringer {\it et al.} (Particle Data Group),
 Phys. Rev. D {\bf 86}, 010001 (2012).

\bibitem{CLEO_dbar} D.M.~Asner {\etal} (CLEO Collaboration),
Phys. Rev. D {\bf 75}, 012009 (2007), see also
H.~Albrecht {\etal} (ARGUS Collaboration), 
Phys. Lett. B {\bf 236}, 102 (1990).

\bibitem{conj} The inclusion of charge-conjugate modes
is implied unless explicitly stated otherwise.

\bibitem{KEKB}
S.~Kurokawa and E.~Kikutani, Nucl. Instr. and. Meth. A {\bf 499}, 1 (2003),
and other papers included in this volume.

\bibitem{Belle}
A.~Abashian {\it et al.} (Belle Collaboration),
Nucl. Instr. and Meth. A {\bf 479}, 117 (2002) and
Y.~Ushiroda (Belle SVD2 Group),
Nucl. Instr. and Meth. A {\bf 511}, 6 (2003).

\bibitem{pythia} T.~Sj\"{o}strand, S.~Mrenna and P.~Skands,
JHEP {\bf 026}, 0605 (2006).

\bibitem{geant3}
R.~Brun~{\etal}, GEANT 3.21, CERN Report DD/EE/84-1 (1984). 

\bibitem{belle_lambda} K.~Abe {\it et al.} (Belle Collaboration),
Phys. Rev. D {\bf 65}, 091103 (2002).

\bibitem{pideff_lppi} These pid requirements correspond
to a $p$ ($\pi $) efficiency of 86\% (93\%) and 
$h\rt p$ ($K\rt \pi$) misidentification probability less
than 1\% (7.5\%).

\bibitem{ARGUS}  H.~Albrecht~{\etal} (ARGUS Collaboration),
Phys. Lett. B {\bf 241}, 278 (1990). We use:
$f(x)=x\sqrt{(x/m_{0})^2 -1}\, \exp[-a((x/m_{0})^2 -1)]$.

\bibitem{nyns} X.L.~Wang {\etal} (Belle Collaboration),
Phys. Rev. D {\bf 84}, 071107(R) (2011).

\bibitem{limits} The sample of $\lm\lm$ plus $\lmb\lmb$
events with $M(\lm\lm)$ below the $m_{\xim}+m_{p}$ threshold
detected in this experiment, as determined from the 
fits to the two dimensional $M(p_1\pim_1)$ {\it vs.}
$M(p_2\pim_2)$ histograms described in the text, contains
$2.3\times 10^{3}$ events and is nearly two orders of
magnitude larger than the 28 $\lm\lm$ events
with invariant mass below $m_{\xim}+m_{p}$ used
in KEK experiment E522.


%


\end{thebibliography}
\end{document}